\documentclass[prl, twocolumn, superscriptaddress, longbibliography]{revtex4-2}
\usepackage{graphicx,epsfig} 
\usepackage{amsmath}
\usepackage{amsfonts}
\usepackage{cancel}
\usepackage[colorlinks=true, allcolors=blue]{hyperref}
 \usepackage{soul}
\newcommand{\sgn}{\operatorname{sgn}}

\newcommand{\ve}[1]{\boldsymbol{#1}}

\begin{document}
\title{Altermagnetic Anomalous Hall Effect Emerging from Electronic Correlations}

\author{Toshihiro Sato}
\affiliation{\mbox{Institute for Theoretical Solid State Physics, IFW Dresden, 01069 Dresden, Germany}}
\affiliation{\mbox{W\"urzburg-Dresden Cluster of Excellence ct.qmat, Germany}}

\author{Sonia Haddad}
\affiliation{Laboratoire de Physique de la Mati\`ere Condens\'ee, Facult\'e  des Sciences de Tunis, Universit\'e Tunis El Manar, Campus Universitaire 1060 Tunis, Tunisia}
\affiliation{\mbox{Institute for Theoretical Solid State Physics, IFW Dresden, 01069 Dresden, Germany}}
\affiliation{\mbox{Max-Planck-Institut f\"ur Physik komplexer Systeme, 01187 Dresden, Germany}}

\author{Ion Cosma Fulga}
\affiliation{\mbox{Institute for Theoretical Solid State Physics, IFW Dresden, 01069 Dresden, Germany}}
\affiliation{\mbox{W\"urzburg-Dresden Cluster of Excellence ct.qmat, Germany}}

\author{Fakher F. Assaad}
\affiliation{\mbox{Institut f\"ur Theoretische Physik und Astrophysik, Universit\"at W\"urzburg, 97074 W\"urzburg, Germany}}
\affiliation{\mbox{W\"urzburg-Dresden Cluster of Excellence ct.qmat, Germany}}

\author{Jeroen van den Brink}
\affiliation{\mbox{Institute for Theoretical Solid State Physics, IFW Dresden, 01069 Dresden, Germany}}
\affiliation{\mbox{W\"urzburg-Dresden Cluster of Excellence ct.qmat, Germany}}

\date{\today}

\begin{abstract}
While altermagnetic materials are characterized by a vanishing net magnetic moment, their symmetry in principle allows for the existence of an anomalous Hall effect. 
Here, we introduce a model with altermagnetism in which the emergence of an anomalous Hall effect is driven by interactions.
This model is grounded in a modified Kane-Mele framework with antiferromagnetic spin-spin correlations.
Quantum Monte Carlo simulations show that the system undergoes a finite temperature phase transition governed by a primary antiferromagnetic order parameter accompanied by a secondary one of Haldane type.
The emergence of both orders turns the metallic state of the system, away from half-filling, to an altermagnet with a finite anomalous Hall conductivity. A mean field ansatz corroborates these results, which pave the way into the study of correlation induced altermagnets with finite Berry curvature.
\end{abstract}

\maketitle
{\it Introduction ---} Ferromagnetic conductors are generally endowed with an observable anomalous Hall effect (AHE), where an electric current perpendicular to the magnetization gives rise to a transversal Hall potential~\cite{Macdo-AHE}. Here, the magnetization of the ferromagnet takes over the role of an external magnetic field that is the root cause of the standard Hall response in time-reversal invariant (i.e. nonmagnetic) conductors. On this basis one might expect that in a fully compensated antiferromagnet the lack of a net magnetization forces the AHE response to vanish. Interestingly, a symmetry analysis shows that this is not necessarily the case --- only time-reversal symmetry breaking by itself is a sufficient condition to allow for an AHE, also in the absence of a net ferromagnetic moment. However, the combination of time-reversal {\it and} translation symmetry, which characterizes most antiferromagnetic materials, implies a vanishing AHE.  Altermagnets comprise the class of compensated collinear antiferromagnets without this combined symmetry~\cite{Libor20,Hayami,Yuan20,Libor22,Mazin21,Macdo22,Libor22-2} and a large set of altermagnetic materials have been identified by first principles electronic structure calculations~\cite{Yaqian, AM-2D}. They are characterized by a fully compensated magnetic order and therefore a zero net magnetic moment, but their symmetry properties reveal that this type of compensated magnetic order may induce an AHE~\cite{Libor20, Libor21, RuO-22, Betancourt23, Bai22, AM-obs}.

In spite of the clear ground-state symmetry considerations, so far no interacting models have been proposed in which an altermagnetic AHE emerges. The latter requires a time-reversal invariant Hamiltonian in which time-reversal symmetry (TRS) is spontaneously broken with zero total moment and still finite anomalous Hall conductivity.
Here we show that precisely such an altermagnetic order emerges in a modified Kane-Mele model with broken inversion symmetry and antiferromagnetic spin-spin interactions. Quantum Monte Carlo calculations reveal that, at finite temperature, the primary antiferromagnetic (AFM) order parameter gives rise to a secondary altermagnetic one, inducing a finite AHE. The occurrence of these two order parameters is in full agreement with the recently developed Landau theory of altermagnetism~\cite{McClarty24}. 
The smoking gun of this emergent altermagnetic phase is a spin-split electronic band structure with an anomalous Hall conductivity that can be tuned by doping. 

\begin{figure}
\centering
\centerline{\includegraphics[width=0.5\textwidth]{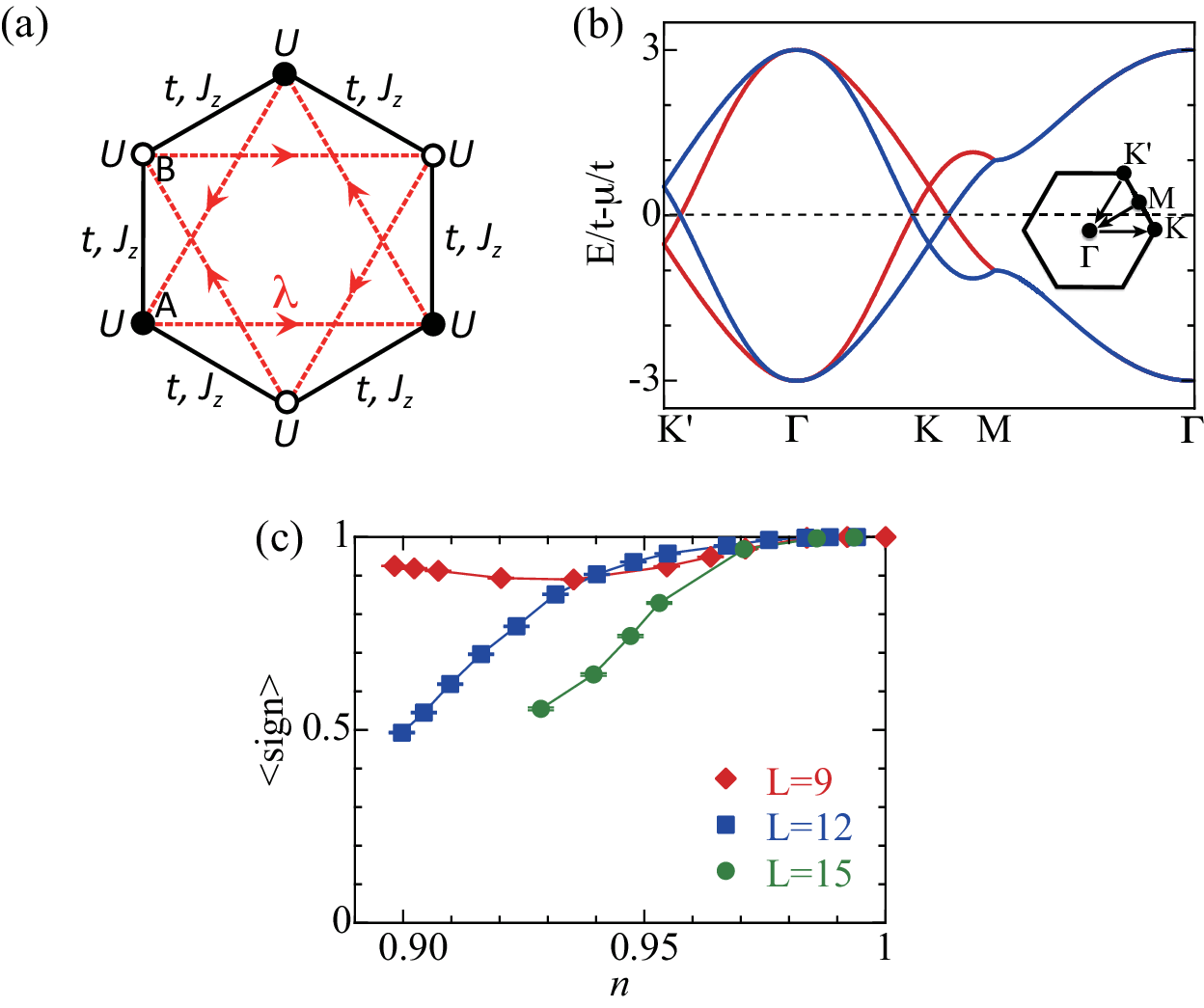}}
\caption{\label{fig:model} (a) The model of fermions on the honeycomb lattice with hopping ($t$, $\lambda$) and interaction ($J_z$, $U$) terms in Eq.~(\ref{Htot}).
(b) Noninteracting band structure at $\lambda=0.1$ for the half-filled case. The red (blue) line corresponds to the spin-up (spin-down) component. Inset: the first Brillouin zone is depicted and the black lines indicate the scans considered here.
(c)  Quantum Monte  Carlo average sign,  $\langle \text{sign} \rangle$,  as a function of $L$ for various electron densities $n$ at temperature $T=1/20$.}
\end{figure}

{\it Interacting altermagnetic Kane-Mele model ---} We start with a modified Kane-Mele (KM) model on a honeycomb lattice with a unit cell containing two sites denoted by A and B. In contrast to the canonical KM model, the sign of the complex phase of the hopping integrals between next nearest neighboring (NNN) sites is opposite on the two sublattices, as in Ref.~\cite{Franz}. 
The corresponding Hamiltonian is
\begin{eqnarray}
\hat{H}_0=-t \sum_{\langle i,j\rangle,s }\hat{c}_{i,s}^{\dagger} \hat{c}^{}_{j,s} - \lambda \sum_{\langle\langle i,j\rangle\rangle ,s} e^{i s \Phi_{i,j}} \hat{c}_{i,s}^{\dagger} \hat{c}^{}_{j,s}
-\mu \sum_{i,s}\hat{n}_{i,s},  \nonumber \\
\label{mKM}
\end{eqnarray}
where $\hat{c}_{i,s}$ is the annihilation operator of an electron of spin $s$ on a honeycomb lattice site, $\hat{n}_{i,s} \equiv \hat{c}_{i,s}^\dagger \hat{c}^{\phantom{\dagger}}_{i,s}$, and $\mu$ is the chemical potential. $t$ and $\lambda$ are the hopping integrals between the nearest neighboring (NN) and NNN sites, respectively. $ \Phi_{i,j}= \pm \pi/2$ is the complex phase gained by an electron during a NNN hopping process according to
the pattern shown in Fig.~\ref{fig:model} (a). 
Effectively the $\lambda$ term introduces a pseudoscalar potential that offsets the Dirac cones for the two spin components, thus generating anisotropic spin-split electronic bands [see Fig.~\ref{fig:model} (b)].
For any finite value of $\lambda$, the SU(2) total spin symmetry is reduced to a U(1) one, and the inversion symmetry is broken, while TRS remains unbroken.  
Note that  the canonical KM model maintains both time-reversal and inversion symmetries, resulting in degenerate spin states without spin splitting.
For an altermagnetic state to emerge, spin-spin interactions must maintain zero net magnetization while they spontaneously break TRS.
Accordingly we consider an AFM interaction term, ensuring that each spin on one sublattice is coupled exclusively to spins on the opposing sublattice.
The total Hamiltonian, taking into account an appropriate interaction that realizes the aforementioned physical characteristics, reads
\begin{eqnarray}
\hat{H}=\hat{H}_0-\frac{J_z}{2}\sum_{\langle i,j\rangle} \left(\hat{S}^z_{i}-
\hat{S}^z_{j} \right)^2,
\label{Htot}
\end{eqnarray}
where $\hat{S}^{z}_{i}=\frac{1}{2}\sum_{\sigma\sigma'} \hat{c}_{i\sigma}^{\dagger} {\sigma}^z_{\sigma\sigma'} \hat{c}^{}_{i\sigma'}$
is the fermion spin operator and $\boldsymbol{\sigma}$ corresponds to the vector of Pauli spin-1/2 matrices.
We consider $J_z >0$ such  that  the $\hat{S}^z_{i} \hat{S}^z_{j}$ term harbors the potential for the development of long-range AFM order in the  spin-$z$ direction.
In addition an effective on-site Hubbard repulsion term $U\sum_{i} \hat{n}_{i \uparrow} \hat{n}_{i \downarrow}$ with $U=\frac{3J_z}{4}$  is generated, which preempts any local pairing instability.

{\it Symmetry analysis ---} To establish that the effective Landau theory of our model is of altermagnetic nature~\cite{Paul}, we consider the continuum  limit where the low energy effective Hamiltonian is 
\begin{equation}
  H^{\text{eff}}_0    =  \sum_{k} \ve{\Psi}^{\dagger}_{k}   \left(  k_x \tau^x \mu^z   -  k_y   \tau^y \right)  \ve{\Psi}^{\phantom\dagger}_{k}    +    \lambda  \sum_{k}  
  \ve{\Psi}^{\dagger}_{k}    \mu^z \sigma^z \ve{\Psi}^{\phantom\dagger}_{k}. \nonumber
\end{equation}
Here $ \ve{\Psi}^{\dagger}_{k}  = \Psi^{\dagger}_{k,\tau,\mu,\sigma} $   and $ \tau,\mu,\sigma$  account  for  the  sublattice,  valley  and   spin  indices, respectively,  on  which  the    Pauli  matrices    $ \tau^{\alpha},\mu^{\alpha},\sigma^{\alpha}$   act.    The  first  term  corresponds  to   the  Dirac  Hamiltonian of  graphene  with  unit   Fermi  velocity   and the second  term  to the   NNN hopping of the  lattice  model.    
Importantly,  we  note   that  this term  differs from the    
quantum spin Hall  mass, which takes the form   $ \sum_{k}  \ve{\Psi}^{\dagger}_{k}    \mu^z  \tau^z \sigma^z \ve{\Psi}^{\phantom\dagger}_{k} $~\cite{KaneMele05}.    It  does  not   open  a  gap in the spectrum, but   shifts    the  valleys in energy  by  $  \lambda \mu^z \sigma^z $, as is seen in Fig.~\ref{fig:model}(b).
The Hamiltonian  is  invariant  under   global U(1)  spin rotations  around  the  $z$ axis, while discrete  symmetries  include    time  reversal: 
$T^{-1} \alpha  \Psi^{\dagger}_{k} T^{}  =  \overline{\alpha}  \Psi^{\dagger}_{-k} \mu^x i \sigma^y$, where $\bar{\alpha}$ denotes the complex conjugation of the $\alpha$.
Importantly,  the    $\lambda$ term   breaks   inversion   symmetry: 
$
I^{-1} \alpha  \Psi^{\dagger}_{k} I  =    \alpha  \Psi^{\dagger}_{-k} \tau^x  \mu^x.
$
The  interaction  term  generates  
\begin{equation}
	H_I   =  - J_z  \int d^2{\ve{x}}     \left(   \ve{\Psi}^{\dagger}(\ve{x})  \sigma^z   \tau^z   \ve{\Psi}^{\phantom\dagger}(\ve{x})  \right)^2,  
\end{equation}
which does  not  lower the  symmetry of  the model,  and  clearly  promotes    antiferromagnetism.  
To  proceed, we  now  consider   a symmetry broken  antiferromagnetic  state in  the  spin-$z$  direction   described by the collinear  N\'eel order  parameter
\begin{equation}
	   N_{\parallel}( \ve{x} )   =     \left< \ve{\Psi}^{\dagger}(\ve{x})  \sigma^z   \tau^z   \ve{\Psi}^{\phantom\dagger}(\ve{x})   \right> .
\end{equation}
This  order   parameter    solely   breaks   TRS such  that a  Ginzburg-Landau  theory     accounting for  this   state can   include     even  powers  of   $N(\ve{x})$.     However, terms  of  the  form   $N(\ve{x}) M(\ve{x})  $   are  also  allowed  in  the 
Ginzburg-Landau functional      provided  that    $M(\ve{x}) $ is odd  under  time  reversal        and  that
$N(\ve{x}) M(\ve{x})  $  shares the  same  symmetries  as  the  Hamiltonian.  
This requirement is fulfilled by the  Haldane  mass \cite{Haldane98}
\begin{equation}   
	 M_{H} ( \ve{x} )  =   \left< \ve{\Psi}^{\dagger}(\ve{x})  \tau^z   \mu^z  \ve{\Psi}^{\phantom\dagger}(\ve{x})   \right>   
\end{equation} 
which indeed  is  odd  under time reversal and generates the AHE.  Since  at  $\lambda \ne 0$  the  Hamiltonian  does not  enjoy  inversion  symmetry,       $N_{\parallel}(\ve{x}) M_{H}(\ve{x})  $   is   allowed  in  the  Ginzburg-Landau  
theory.  As  a  consequence,  as  soon as  $N_{\parallel}(\ve{x})$  acquires  a nontrivial  expectation value,  the  Haldane  mass  is  generated  as a  secondary order parameter~\cite{Paul}.    
We  note  that  at  half-band filling,   characterized  by  the  particle-hole  symmetry    $ P^{-1} \alpha \ve{\Psi}^{\dagger}_{\ve{k}}   P =  \overline{\alpha} \ve{\Psi}_{\ve{k}}^{T}  \tau^z \sigma^x   $,  the    linear  coupling  between  the   two  order parameters  is  forbidden   since 
$N_{\parallel}$ is  even  and  $M_H$  is  odd  under  this  symmetry.   
 It is worth stressing that in the case of an in-plane magnetic ordering, the Ginzburg-Landau formulation will be rephrased in terms of a two-component primary order parameter $\ve{N}_{\perp}( \ve{x} )   =     \left< \ve{\Psi}^{\dagger}(\ve{x})     \tau^z \left(\sigma^x,\sigma^y\right)  \ve{\Psi}^{\phantom\dagger}(\ve{x})   \right>$, which  breaks  $U(1)$  spin  symmetry  as  well as  TRS. As a consequence,  it cannot be linearly coupled to the scalar secondary order parameter $M_H(\ve{x}) $, forbidding by symmetry the emergence of $M_H(\ve{x}) $ and the subsequent altermagnetic properties.

On the lattice, it is clear that the model harbors magnetism beyond simple AFM order as the collinear AFM ordered spin state on the two sublattices cannot be connected by a translation or inversion symmetry combined with time reversal, due to the specific phase patterns in the NNN hopping Hamiltonian, as is expected for an altermagnet~\cite{Libor22}.
Indeed, we will show that this results in a finite total Berry curvature away from half-filling, in sync with the finite Haldane mass in the continuum description.

\begin{figure}
\centering
\centerline{\includegraphics[width=0.5\textwidth]{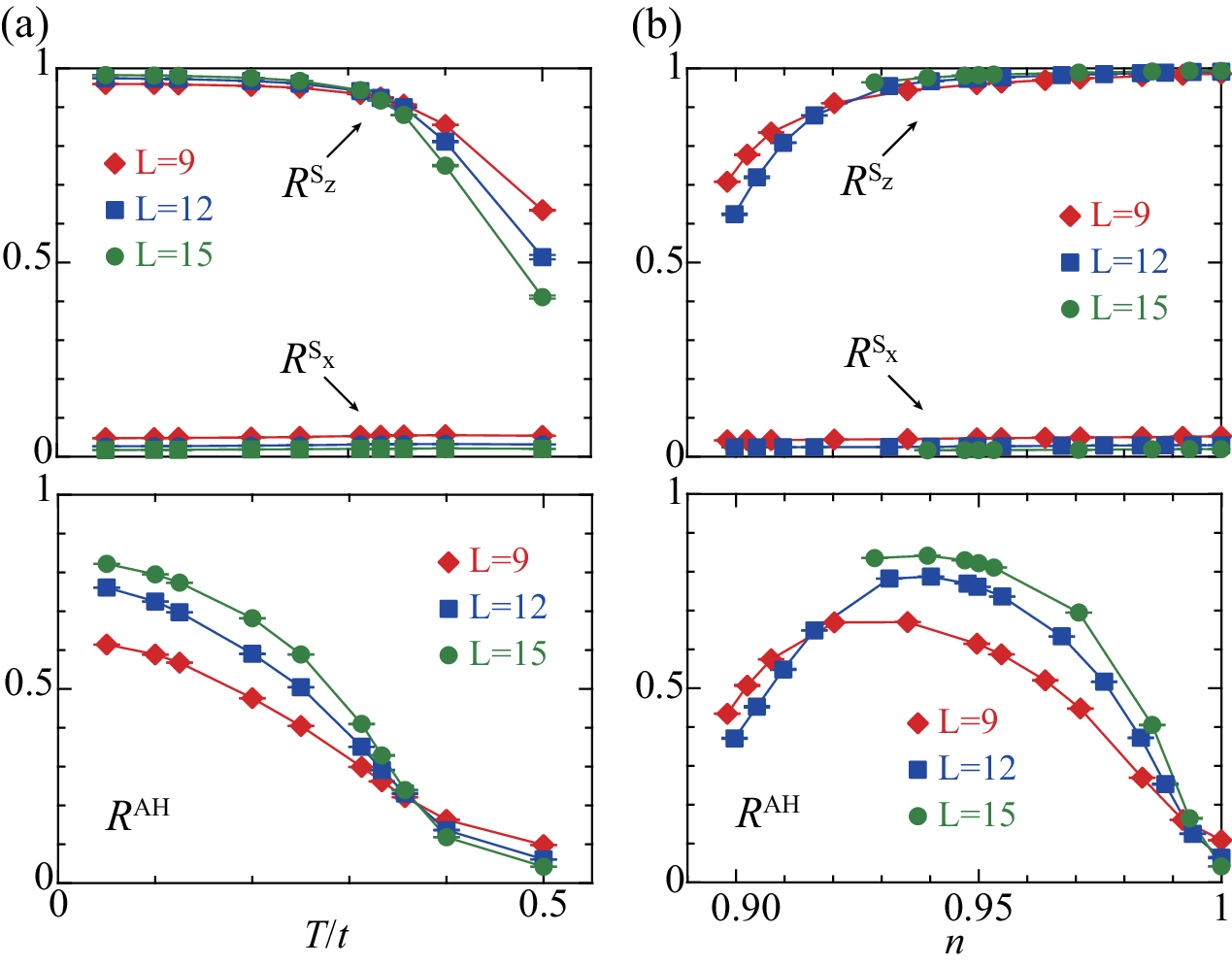}}
\caption{\label{fig:CR} (a) Temperature dependence of correlation ratios for $z$-AFM and $x$-AFM orders (top panel) and AH order (bottom panel) at electron density $n=0.95$. Here, $L$ is the system size, $\lambda=0.1$ and $J_z=2$ in units of $t$.
(b) Same as (a) for varying electron density $n$ at temperature $T=1/20$. 
}
\end{figure}

{\it Quantum Monte Carlo results ---} The Hamiltonian~(\ref{Htot}) was simulated using the ALF (Algorithms for Lattice Fermions) implementation \cite{ALF_v1,ALF_v2} of the grand-canonical, finite-temperature, auxiliary-field quantum Monte Carlo method~\cite{Blankenbecler81, White89,  Assaad08_rev}.
In fact the interaction part of Eq.~(\ref{Htot}) in terms of perfect squares can be implemented in the ALF implementation.
Results were obtained on lattices with $L\times L$ unit cells ($2L^2$ sites) and periodic boundary conditions. 
Henceforth, we use $t=1$ as the energy unit, set $\lambda=0.1$,  $J_z=2$
and  set the  Trotter imaginary  time  step to   $\Delta\tau=0.1$.
The negative sign problem is absent at half-filling  since   in  a  Bogoliubov  basis,   
$(\hat{\gamma}^{\dagger}_{{\mathbf i},\uparrow}, \hat{\gamma}^{\dagger}_{{\mathbf i},\downarrow} )   = 
 (\hat{c}^{\dagger}_{{\mathbf i},\uparrow}, (-1)^{{\mathbf i}}\hat{c}^{}_{{\mathbf i},\downarrow} )  $, 
and  after   decoupling  the  perfect  square  term  with a   Hubbard-Stratonovitch transformation, 
time-reversal and  U(1)  charge  symmetries  are present for each field configuration \cite{Wu04}.   
Doping   breaks  this  symmetry    and  the sign problem  sets  in. 
Nevertheless,  for our  specific  implementation  it   turns  out to  be mild [see Fig.~\ref{fig:model} (c)] such that   {\it large}  lattices  and  {\it low}  temperatures
can  be  reached.  In particular, we  are  able  to  reveal an   altermagnetic phase,  as  shown in Fig.~\ref{fig:CR}.
In fact, this was achieved by viewing the sign problem as an optimization problem over the space of possible path integral formulations.

To verify the above low-energy theory, we measure equal-time correlation functions 
\begin{equation}
C^{O}(\mathbf q)\equiv\frac{1}{L^2}\sum_{\mathbf r, \mathbf r'}\langle  \hat{O}^{O}_{\mathbf r} \hat{O}_{\mathbf r'}^{O}\rangle e^{i \mathbf q \cdot (\mathbf r-\mathbf r')}
\label{eq:Cq}
\end{equation}
of fermion spin  $  \hat{O}_{\mathbf r}^{S_z(S_{xy})} = \hat{S}^{z(xy)}_{\mathbf{ r},A} - \hat{S}^{z(xy)}_{\mathbf{r},B} $ with 
$\hat{S}^{s}_{\mathbf{ r}}= \frac{1}{2} \hat{c}^{\dagger}_{\mathbf{ r}} \sigma^{s} \hat{c}^{\phantom \dagger}_{\mathbf{ r}}$, 
and the operator corresponding to the Haldane mass,
$\hat{O}^{\text{AH}}_{\boldsymbol{r}} =
\sum_{\langle\langle \boldsymbol{\delta},\boldsymbol{\delta'} \rangle\rangle, \sigma}
e^{i \Phi_{\boldsymbol{r} + \boldsymbol{\delta}, \boldsymbol{r} + \boldsymbol{\delta}' }}
 \hat{c}^{\dagger}_{\boldsymbol{r}+ \boldsymbol{\delta},\sigma}
\hat{c}^{}_{\boldsymbol{r}+ \boldsymbol{\delta'},\sigma} $.
Here, $\ve{r}$ specifies a unit cell,  or  hexagon, and   $\langle\langle \boldsymbol{\delta},\boldsymbol{\delta'} \rangle\rangle$ runs  over  the  NNN sites  of this hexagon.
After computing the correlation functions, we extracted the renormalization-group invariant correlation ratios \cite{Binder1981,Pujari16}
\begin{equation}
R^{O}=1-\frac{C^{O}({\mathbf  q_0}+\delta {\mathbf   q}) }{C^{O}(\mathbf  q_0)}.
\label{eq:R}
\end{equation}
Here ${\mathbf  q_0}$ is the ordering wave vector and ${\ve q}_0 + \delta {\ve q}$ the longest wave-length fluctuation of the ordered state for a given lattice size.
Long-range AFM and anomalous Hall (AH) orders here imply a divergence of the corresponding correlation functions 
$C^{O}(\mathbf  q_0=0)$.
Accordingly, 
$R^{O} \to 1$
for $L\to\infty$ in the ordered phase, whereas 
$R^{O} \to 0$
 in the disordered phase.  
At the critical point, 
$R^{O}$
 becomes scale-invariant for sufficiently large system size $L$, leading to a crossing of results for different $L$. 

We find the altermagnetic phase in the range of parameters, specifically temperatures $T$ and electron densities $n$, shown in Fig.~\ref{fig:CR}.
Figure~\ref{fig:CR} (a) shows the results as a function of $T$ at $n=0.95$.
The onset of the AFM in spin-$z$ ($x$) direction, termed $z$-AFM ($x$-AFM), is detected from the crossing of $R^{S_z} (R^{S_x})$, whereas the onset of the AH order can be detected from $R^\text{AH}$.  
As  the  temperature decreases, we observe the onset of the $z$-AFM order.   We  note  that  $z$-AFM   only  breaks  time-reversal  
symmetry  such that ordering  at  finite  temperature is  allowed. 
A key feature is that its appearance coincides with the onset of the AH order.
Indeed, the data for $R^\text{AH}$ are consistent with a finite-temperature phase transition to the $z$-AFM phase.
Furthermore, within the range of parameters we investigated, we do not observe the emergence of the $x$-AFM order.
Results as a function of $n$ for $T=1/20$ are presented in Fig.~\ref{fig:CR} (b).
The consequence of the simultaneous emergence of both $z$-AFM and AH orders persists against changes in electron density away from half filling.
At   half-filling     the   $z$-AFM  ordering  persists   while  AH correlations are  suppressed.    This agrees with our finding of a vanishing linear coupling between the AH and $z$-AFM  order in the  Ginzburg-Landau  functional when particle-hole  symmetry is present. 
As $n$ is further decreased, the correlation ratios do not provide clear evidence of both these orders, transitioning to the disordered phase.

We  conclude  this  section by noting  that  we  have  checked   that no  ferromagnetic order   emerges,  such  that  the  state  that  we  observe is 
a  compensated  collinear  antiferromagnet. 

{\it AHE in mean field approximation ---} Within a mean field (MF) approximation the AHE that emerges in the interacting  modified KM model
  can be accessed directly. Using the spin-dependent sublattice basis $\left(\psi_{A\uparrow},\psi_{B\uparrow},\psi_{A\downarrow},\psi_{B\downarrow}\right)$, the corresponding Hamiltonian, given by  Eq.~(\ref{Htot}), reduces to
\begin{eqnarray}
H^{\text{MF}}({\mathbf{k}})=(\epsilon-\mu)\sigma^0\tau^0+a_{\mathbf{k}} \tau^0 \sigma^z+  \mathbf{h}({\mathbf{k}})\cdot\ve{\tau},
\label{H-MF}
\end{eqnarray}
where $h(\mathbf{k})=(b_{\mathbf{k}}{ \sigma^0},c_{\mathbf{k}}{ \sigma^0},h_z)$, $h_z=- \Delta M^{\text{AFM}}_z\sigma^z $,  
$\Delta=\frac32 J_z+\frac 34 |J_z|$, $\mu$ is the chemical potential, $\epsilon=\frac38 |J_z|(n-1)$, $n=n_a=n_b$ is the electron density on site A or B, with
$n_l=\langle \hat{n}_{l\uparrow}\rangle+\langle \hat{n}_{l\downarrow}\rangle$, $l=a,\,b$ being the sublattice index. The antiferromagnetic $M^{\text{AFM}}_z$ order parameter along the $z$ direction is expressed in terms of the on-site magnetization as $M^{\text{AFM}}_z=\frac12 \left(m^a_z -m^b_z \right)$, where
$m^l_z=\frac 12 \left(\langle \hat{n}_{l\uparrow}\rangle-\langle \hat{n}_{l\downarrow}\rangle \right)$. The hopping terms $ a_{\mathbf{k}}$, $ b_{\mathbf{k}}$ and $ c_{\mathbf{k}}$ are given by
$a_{\mathbf{k}}={-}2\lambda \sum_{i=1}^3 \sin\left( \mathbf{k}\cdot \mathbf{a}_i\right)$, 
$b_{\mathbf{k}}={-}t\sum_{i=1}^3 \cos( \mathbf{k}\cdot \boldsymbol{\delta}_i)$, 
$c_{\mathbf{k}}={-}t\sum_{i=1}^3 \sin(\mathbf{k}\cdot \boldsymbol{\delta}_i)$.
$\boldsymbol{\delta}_i$ and $\mathbf{a}_i$ are, respectively, the vectors connecting NN  and NNN sites;
$\mathbf{a}_1=\sqrt{3}a(-\frac{1}{2},\frac {\sqrt{3}}{2})$, $\mathbf{a}_2=\sqrt{3}a(-\frac{1}{2},-\frac{{\sqrt3}}{2})$,  $\mathbf{a}_3=\sqrt{3}a(1,0)$, $
\boldsymbol{\delta}_1=a(\frac{\sqrt{3}}{2},\frac{1}{2})$, $\boldsymbol{\delta}_2=a(-\frac{\sqrt{3}}{2},\frac{1}{2})$, $
\boldsymbol{\delta}_3=a(0,-1)$, where $a$ is the distance between NN sites.
$\tau^0$ ($\sigma^0$) is the $2\times2$ identity matrix in the sublattice (spin) space. 

The spectrum of the mean field Hamiltonian (\ref{H-MF}) is given by
\begin{eqnarray}
E^{\text{MF}}_{{\nu},\sigma}({\mathbf{k}})=\epsilon-\mu+ a_{\mathbf{k}} \sigma+{\nu} \sqrt{b^2_{\mathbf{k}}+c^2_{\mathbf{k}}+\left(\Delta M^{\text{AFM}}_z\right)^2},\nonumber
\end{eqnarray}
where 
${ \nu}=\pm$ ($\sigma=\pm$) 
is the band (spin) index.
The spin-dependent mass term $h_z$ in Eq.~(\ref{H-MF}) breaks TRS  and a nonvanishing Berry curvature (BC) is expected away from half-filling. 
For a given spin projection $\sigma$, the BC of a band 
${ \nu}$
can be expressed as~\cite{Qi08a,Fred}
\begin{eqnarray}
\Omega^{{ \nu}}_z(\sigma,\mathbf{k})=-\frac{{\nu}}{2|\mathbf{h}({\mathbf{k}})|^3} 
\mathbf{h}(\mathbf{k})\cdot\left[ \partial_{k_x} \mathbf{h}(\mathbf{k})\times \partial_{k_y}\mathbf{h}(\mathbf{k})\right] \nonumber
\end{eqnarray}
and the AH conductivity~\cite{Macdo22} is given by the integral
\begin{eqnarray}
\sigma_{\rm Hall}=-\frac {e^2}{\hbar} \frac{1}{(2\pi)^2}\sum_{\sigma,{ \nu}}\int_{BZ}d\mathbf{k}\;
\Omega^{{\nu}}_z(\sigma,\mathbf{k}) f_{{\nu}}(\mathbf{k},\mu),\nonumber
\end{eqnarray}
where $h(\mathbf{k})=(b_{\mathbf{k}},c_{\mathbf{k}},- \Delta M^{\text{AFM}}_z\sigma)$ and 
$f_{{\nu}}(\mathbf{k},\mu)$ 
is the Fermi-Dirac distribution function ~\cite{BZ-square}.\

\begin{figure} 
\centering
\includegraphics[width=0.65\columnwidth]{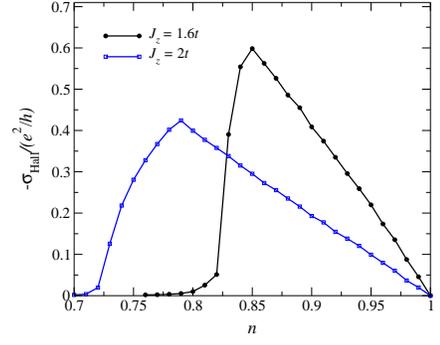}
\caption{
Anomalous Hall conductivity $\sigma_{\rm Hall}$ in units of $e^2/h$
as a function of the electron density $n$ for $J_z=1.6t$ and $J_z=2t$ using the mean field parameters for $n$ and $M^{\text{AFM}}_z$. 
}
\label{Fig_sigma}
\end{figure}

$\sigma_{\rm Hall}$ can be computed using the values of the chemical potential and the AFM magnetization $M^{\text{AFM}}_z$ extracted from the mean field calculations. The results are depicted in Fig.~\ref{Fig_sigma} showing $\sigma_{\rm Hall}$ as function of the electronic density $n$. 

The occurrence of a nonvanishing AH conductivity can be understood from the spin-valley dependence of the BC and the offset of the Dirac points induced by the modified KM term $a_{\mathbf{k}} \sigma^z\tau^0$ of Eq.~(\ref{H-MF}), which shifts oppositely the spin-split bands. To gain further insight into these features, let us focus on the states $\mathbf{k}=\mu_z\mathbf{K}+\mathbf{q}$, ($\mathbf{q}\ll |\mathbf{K}|$) in the vicinity of the Dirac points $\mu_z\mathbf{K}$, where $\mu_z=\pm$ is the valley index. These states contribute strongly to the BC, which reduces, around $\mu_z\mathbf{K}$, to 
\begin{equation}\label{eq:Omegaz}
\Omega^{{ \nu}}_z(\sigma,\mu_z,\mathbf{q})=\sgn \frac{(\hbar v_F)^2 \Delta  M^{\text{AFM}}_z}{2\left[(\hbar v_F \mathbf{q})^2+\left(\Delta  M^{\text{AFM}}_z\right)^2\right]^{3/2}},
\end{equation}
where 
$\sgn=-{\nu}{\mu_z} \sigma $ is the sign of the BC and $\hbar v_F=\frac 32 a t$.
Given the spin-dependent offset of the Dirac points, a spin-split subband 
$E^{\text{MF}}_{{\nu},\sigma}({\mathbf{k}})$ [$E^{\text{MF}}_{{\nu},-\sigma}({\mathbf{k}})$]
 of an occupied band ${\nu}$, contributes to the BC by the states around the valley $\mu^z$ ($-\mu^z$).
 Thus, the two contributing spin-dependent BCs have the same sign, resulting in a nonvanishing total BC away from half-filling.
 Data are shown in the Supplemental Material~\cite{supp}.
 We note that in the continuum limit, our modified KM term, $a_{\mathbf{k}}\tau^0\sigma^z$, indeed corresponds to a valley-Zeeman spin-orbit coupling, akin to that considered in transition metal dichalcogenides~\cite{Xiao12,Wang2022}. 
Both systems exhibit effects that induce spin-split band structures and enable spin-valley locking, but their underlying origins differ, as detailed in the Supplemental Material~\cite{supp}.

The doping dependence of the AH conductivity (see Fig.~\ref{Fig_sigma}) reflects the behavior of the secondary order parameter of Haldane type $M_{\rm H}$. 
At half-filling ($n=1$), $\sigma_{\rm Hall}$ vanishes as the bands below the Fermi level are fully occupied, resulting in zero total BC. Decreasing $n$ from half-filling, $|\sigma_{\rm Hall}|$ increases up to a maximum value, at a doping $n_{c}$, where it drops and vanishes below a critical doping value $n_{c0}$. Larger coupling constants $J_z$ give smaller $n_{c0}$. 
For a fixed $J_z$, the increase of $|\sigma_{\rm Hall}|$ can be understood from the BC contribution which is enhanced as the area of the spin-polarized Fermi surface increases (see the Supplemental Material~\cite{supp}). 
The drop of $|\sigma_{\rm Hall}|$ is a consequence of the sharp decrease of $M^{\text{AFM}}_z$ below $n_{c}$, inducing a decrease of the BC.
The doping regime with a nonzero AH conductivity gets wider as  $J_z$ increases, which results from the enhancement of the mass term $M^{\text{AFM}}_z$.

{\it Conclusions and outlook ---} We have shown how electronic interactions induce altermagnetism in a noncentrosymmetric system by stabilizing a primary AFM ordering that hosts a secondary one that directly induces the altermagnetic anomalous Hall effect.
Specifically, our quantum Monte Carlo calculations on the interacting modified Kane-Mele lattice model reveal a finite temperature phase transition into an altermagnetically ordered state, whose physical features are captured by its effective continuum theory.
The interacting model may in principle be implemented in cold atoms in optical lattices that have been proposed to realize altermagnetism~\cite{AM-cold}. 
Time-modulated optical lattices of ultracold atoms allow for complex NNN tunneling, as demonstrated for the Haldane model~\cite{Jotzu:2014aa}, while Hubbard models with AFM order have been realized in ultracold fermions~\cite{Hart:2015aa,Mazurenko:2017aa}. 
This suggests potential for combining these techniques, incorporating both complex NNN tunneling and Hubbard-type models with AFM order.
Interestingly, recently an experimental realization of a non-centrosymmetric 3D altermagnet has been reported in GdAlSi, 
a collinear antiferromagnetic Weyl semimetal~\cite{GdAlSi}. 
This suggests generalization of the lattice model to higher dimensions to determine how electronic correlations generate Berry curvature in altermagnetic semimetals that may as well harbor 3D electronic topology.

\bigskip
\begin{acknowledgments}
{\it Acknowledgments ---}  We  acknowledge  fruitful discussions   with   Paul McClarty, Olena Gomonay, Libor \v{S}mejkal, and {Tom\'a\v{s} Jungwirth}. We  gratefully acknowledge the Gauss Centre for Supercomputing e.V. for funding this project by providing computing time on the GCS Supercomputer SUPERMUC-NG at Leibniz Supercomputing Centre, (project No. pn73xu)
as  well  as  the scientific support and HPC resources provided by the Erlangen National High Performance Computing Center (NHR@FAU) of the Friedrich-Alexander-Universit\"at Erlangen-N\"urnberg (FAU) under the NHR project b133ae. NHR funding is provided by federal and Bavarian state authorities. NHR@FAU hardware is partially funded by the German Research Foundation (DFG) --- 440719683.
FA, TS, JvdB and ICF  thank  the W\"urzburg-Dresden Cluster of Excellence on Complexity and Topology in Quantum Matter ct.qmat (EXC 2147, project-id 390858490). 
S. H. acknowledges the Institute for Theoretical Solid State Physics at IFW (Dresden) and the Max Planck Institute for the Physics of Complex Systems (Dresden) for kind hospitality and financial support.
\end{acknowledgments}


\begin{thebibliography}{43}
\bibitem [{BZ-()}]{BZ-square}%
  \BibitemOpen
  \href@noop {} {}\bibinfo {note} {To carry out the integration over the
  Brillouin zone, it is easier to use the mapping to a square Brillouin zone
  and write the Hamiltonian in terms of the variables $q_x=\mathbf{k}\cdot
  \mathbf{a}_3\; q_y=\mathbf{k}\cdot \mathbf{a}_1$. The Dirac points are at
  $\mathbf{q}_D=\pm(2\pi/3,2\pi/3)$. The $\mathbf{\delta}_i$ vectors can be
  expressed as $\mathbf{\delta}_i=\frac13 \epsilon_{ijk}\left(
  \mathbf{a}_k-\mathbf{a}_j\right)$~\cite{JNPRB}}\BibitemShut {NoStop}%
\end{thebibliography}
%

\clearpage
\section{Supplemental material}
\subsection{Mean-field ansatz}

In the mean filed approximation, the Hamiltonian given by Eq.~(2) in the main text, can be written in the spin-dependent sublattice basis $\left(\psi_{A\uparrow},\psi_{B\uparrow},\psi_{A\downarrow},\psi_{B\downarrow}\right)$ as
{\small
\begin{widetext}
\begin{eqnarray}
H^{\text{MF}}_{x,z}(\mathbf{k})= 
\begin{pmatrix}
a^0_{\mathbf{k}}+a_{\mathbf{k}}+\frac32 J_z m_z^b-\frac 34 |J_z| m_z^a
& \gamma_{\mathbf{k}} & - \frac 34 |J_z| m_x^a& 0\\
\gamma^{\ast}_{\mathbf{k}} & a^0_{\mathbf{k}}+a_{\mathbf{k}}+\frac32 J_z m_z^a-\frac 34 |J_z|m_z^b& 0&- \frac 34 |J_z| m_x^b\\
- \frac 34 |J_z| m_x^a & 0&a^0_{\mathbf{k}} -a_{\mathbf{k}}-\frac32 J_z m_z^b+\frac 34 |J_z| m_z^a & \gamma_{\mathbf{k}}\\
0 & - \frac 34 |J_z| m_x^b& \gamma^{\ast}_{\mathbf{k}}&a^0_{\mathbf{k}} -a_{\mathbf{k}}-\frac32 J_z m_z^a+\frac 34 |J_z| m_z^b\\
\end{pmatrix}\nonumber\\
\label{HMF_supp}
\end{eqnarray} 
\end{widetext}}
where
\begin{eqnarray}
a^0_{\mathbf{k}}&=&\epsilon-\mu+2\lambda \cos \Phi \sum_{i=1}^3 \cos\left( \mathbf{k}\cdot \mathbf{a}_i\right)\nonumber\\
\epsilon&=&\frac38 |J_z|(n-1)\nonumber\\
a_{\mathbf{k}}&=&2 \lambda \sin\Phi \sum_{i=1}^3 \sin\left( \mathbf{k}\cdot \mathbf{a}_i\right)\nonumber\\
\gamma_{\mathbf{k}}&=&{-}t\sum_{i=1}^3 e^{-i \mathbf{k}\cdot \mathbf{\delta}_i}\equiv b_{\mathbf{k}}-i c_{\mathbf{k}}.
\label{hop}
\end{eqnarray}
$\mu$ is the chemical potential, $n=n_a=n_b$ is the electron density of an atom A or B, with
$n_l=\langle \hat{n}_{l\uparrow}\rangle+\langle \hat{n}_{l\downarrow}\rangle$, $l=a,\,b$ being the sublattice index.
$t$ ($\lambda$) is the hopping integral between nearest neighboring (next nearest neighboring) atoms and
$\Phi$ is the phase gained by the electron when hopping to the next nearest neighboring atom. Without loss of generality, we set $\Phi=-\pi/2$~\cite{Franz} since the hopping between next nearest neighboring A atoms is performed anticlockwise [see Fig.~1 in the main text].
$\mathbf{\delta}_i$ are the vectors connecting a B atom to its neighboring A atoms
and $\mathbf{a}_i$ are the vectors relating next nearest neighboring atoms given by
\begin{eqnarray}
\mathbf{a}_1&=&\sqrt{3}a(-\frac 12,\frac {\sqrt{3}}2), \; \mathbf{a}_2=\sqrt{3}a(-\frac 12,-\frac{{\sqrt3}}2),  \, \mathbf{a}_3=\sqrt{3}a(1,0) \nonumber\\
\mathbf{\delta}_1&=&a(\frac{\sqrt{3}}2,\frac12), \; \mathbf{\delta}_2=a(-\frac{\sqrt{3}}2,\frac12),\;
\mathbf{\delta}_3=a(0,-1),\nonumber\\
\label{vec_a}
\end{eqnarray}
where $a$ is the distance between nearest neighboring atoms.
$m^{l}_{j}$ in Eq.~(\ref{HMF_supp}) denote the magnetizations in the $l=a,b$ sublattice along the $j=x,z$ direction: $m^{l}_x=\langle S_{l}^+ \rangle=\langle S_{l}^- \rangle$ and
$m^l_z=\frac 12 \left(\langle \hat{n}_{l\uparrow}\rangle-\langle \hat{n}_{l\downarrow}\rangle \right)$.
We define the ferromagnetic (FM) $M^\text{FM}_j$ and the antiferromagnetic (AFM) $M^{\text{AFM}}_j$ order parameters along the $j$ ($j=x,z$) direction in terms of the on-site magnetizations as $M^{\text{FM}}_j=\frac12 \left(m^a_j + m^b_j \right)$ and $M^{\text{AFM}}_j=\frac12 \left(m^a_j -m^b_j \right)$. The Hamiltonian, given by Eq.~(\ref{HMF_supp}), can then be written as 
\begin{eqnarray}
H^{\text{MF}}_{x,z}({\mathbf{k}})&=& \left(\epsilon-\mu\right)\sigma^0\tau^0
+\left(a_{\mathbf{k}}+\Delta_0M^{FM}_z \right)\tau^0 \sigma^z\nonumber\\
&&+\left(b_{\mathbf{k}} \tau^x +c_{\mathbf{k}} \tau^y \right)\sigma^{0}
-\Delta_x\left( M^{\text{FM}}_x \tau_{0}+M^{\text{AFM}}_x\tau^z\right)\sigma^x\nonumber\\
&&-\Delta\, M^{\text{AFM}}_z\, \tau^z\sigma^z,\nonumber\\
\label{H-MF1_supp}
\end{eqnarray}
where $\Delta_0=\frac32 J_z-\frac 34 |J_z|$, $\Delta_x=\frac 34 |J_z|$, and $\Delta=\frac32 J_z+\frac 34 |J_z|$.

The quantum Monte Carlo and the mean-field results show that the only non-vanishing primary order parameter is the $z$-component AFM phase. The mean-field Hamiltonian, given by Eq.~(\ref{H-MF1_supp}), reduces to [see Eq.~(7) in the main text]\\
\begin{eqnarray}
H^{\text{MF}}_{z}({\mathbf{k}})&=&\left(\epsilon-\mu\right)\sigma^0\tau^0+a_{\mathbf{k}}\tau^0 \sigma^z
+\left(b_{\mathbf{k}} \tau^x +c_{\mathbf{k}} \tau^y \right)\sigma^{0}\nonumber\\
&&-\Delta\, M^{\text{AFM}}_z\, \tau^z\sigma^z\nonumber\\
&\equiv& \left(\epsilon-\mu\right)\sigma^0\tau^0+a_{\mathbf{k}}\tau^0 \sigma^z+\mathbf{h}\left({\mathbf{k}}\right)\cdot \ve{\tau}.
\label{H-MF2_supp}
\end{eqnarray}
Due to the mass term $h_z=-\Delta\, M^{\text{AFM}}_z\, \tau^z\sigma^z$, the system breaks time-reversal symmetry defined as $\mathcal{T}=\tau^0\sigma^y K$, where $K$ denotes complex conjugation.
The mean-field results are depicted in Figs.~\ref{fig:MF} and \ref{Jzdep-n} showing the order parameters, the chemical potential, the band structures, and the corresponding Fermi surfaces.

\subsection{Berry curvature and anomalous Hall conductivity}
The Berry curvature (BC) is calculated in the canonical basis~\cite{Cayssol_2021}, where the Bloch Hamiltonian depends on the atom positions in the unit cell as expressed in the hopping term $\gamma_{\mathbf{k}}$ of Eq.~(\ref{hop}).
$\Omega^{\alpha}_z(\sigma,\mathbf{k})$ reads as:
\begin{widetext}
\begin{eqnarray}
\Omega^{\alpha}_z(\sigma,\mathbf{k})&=&-\alpha \sigma\frac{ \Delta M^{\rm{AFM}}_z }{|\mathbf{h}({\mathbf{k}})|^3} 
 \sqrt{3} (at)^2 \sin\left(\mathbf{k}\cdot \frac{(\mathbf{\delta_2}-\mathbf{\delta_3})}2\right)
\sin\left(\mathbf{k}\cdot \frac{(\mathbf{\delta_3}-\mathbf{\delta_1})}2\right)
\sin\left(\mathbf{k}\cdot \frac{(\mathbf{\delta_1}-\mathbf{\delta_2})}2\right),\nonumber\\
\label{BC-MF3}
\end{eqnarray}
\end{widetext}
where $\alpha=\pm$ and $\sigma=\pm$ are, respectively, the band and the spin indices.

In Fig.~\ref{BC} we plot the BC of the occupied spin sub-bands for a coupling constant $J_z=2t$ and a doping $n=0.79$, at which the anomalous Hall conductivity reaches its maximum. Figure~\ref{BC} shows that, at the opposite valleys, the contributing bands with opposite spin projections have the same sign of the BC, which gives rise to a finite total Berry curvature of the occupied bands.

\begin{figure*}
    \centering
    \begin{tabular}{cc}
            (a) &   \\ 
        \includegraphics[width=0.5\linewidth]{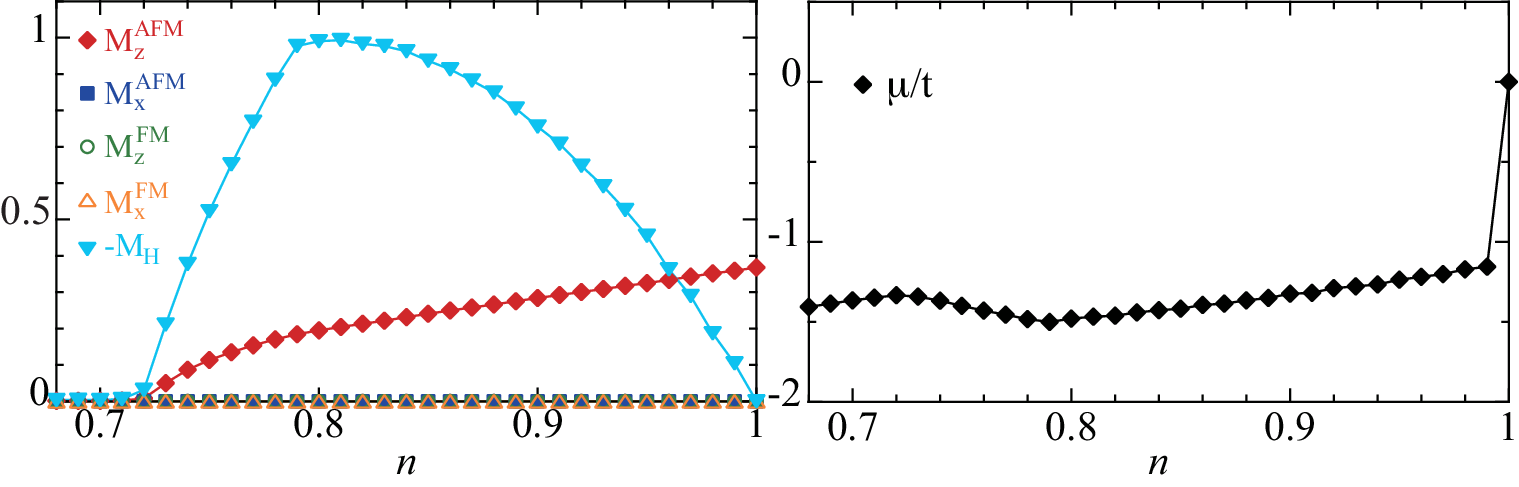} & 
        \includegraphics[width=0.25\linewidth]{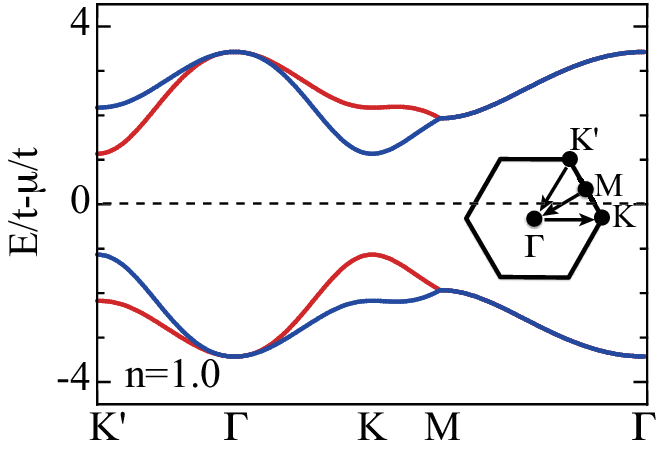} \\ \\
    \end{tabular}
     \begin{tabular}{cccc}
            (b) & & (c) \\ 
        \includegraphics[width=0.25\linewidth]{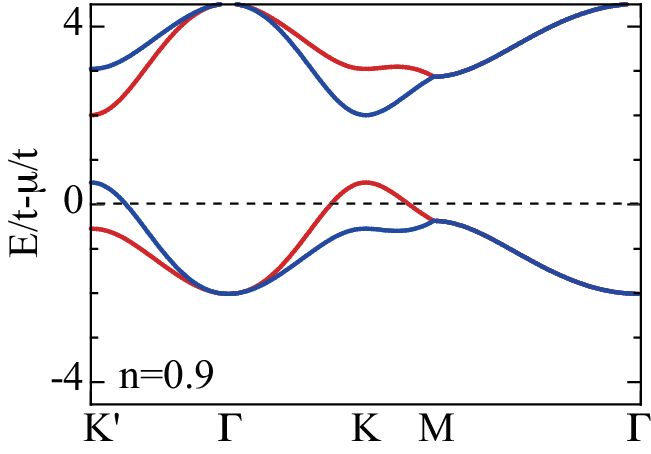} & 
        \includegraphics[width=0.18\linewidth]{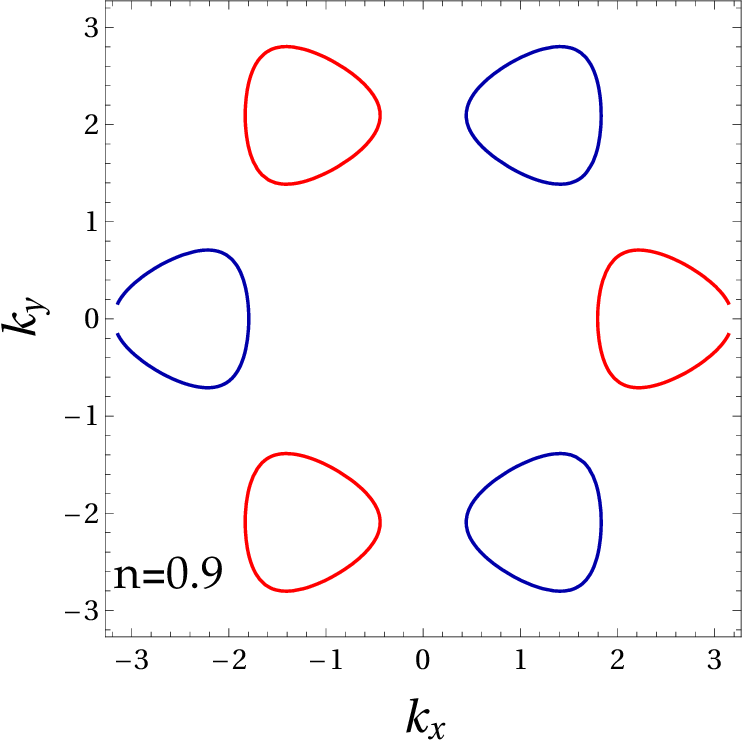}&~~~~
        \includegraphics[width=0.25\linewidth]{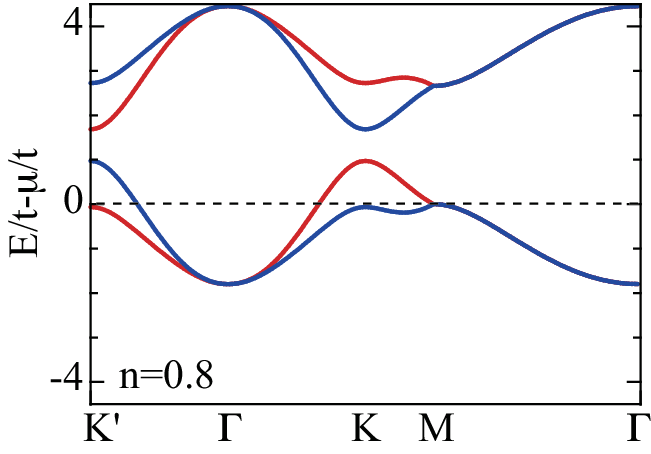} &        
        \includegraphics[width=0.18\linewidth]{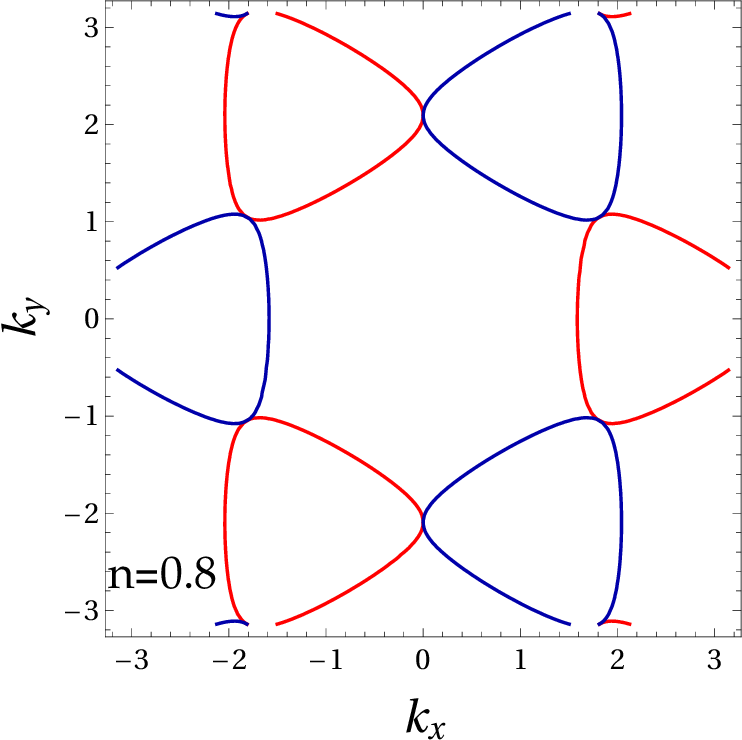} \\ \\
    \end{tabular}
     \begin{tabular}{cccc}
            (d) & & (e) \\ 
        \includegraphics[width=0.25\linewidth]{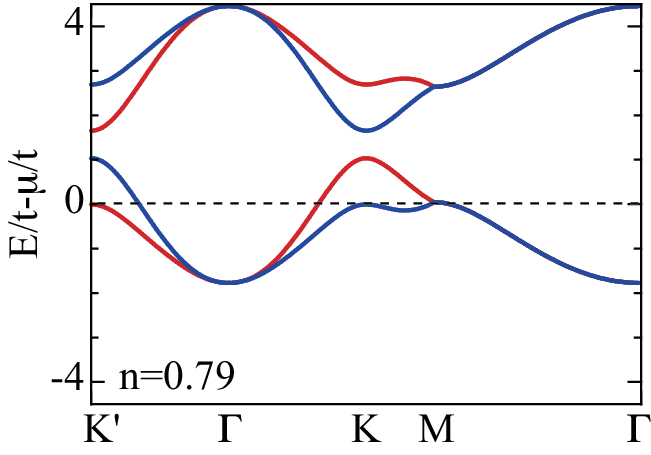} & 
        \includegraphics[width=0.18\linewidth]{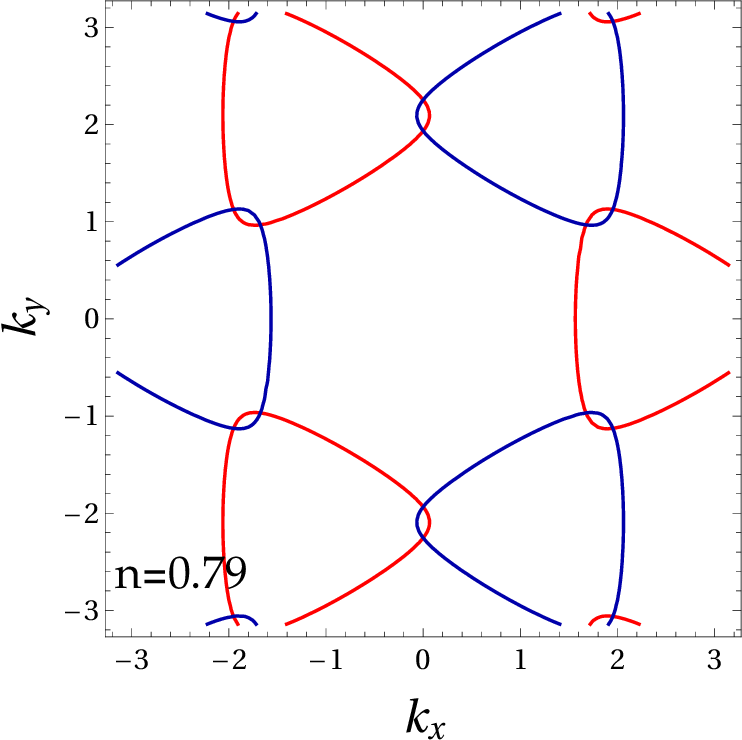}&~~~~
        \includegraphics[width=0.25\linewidth]{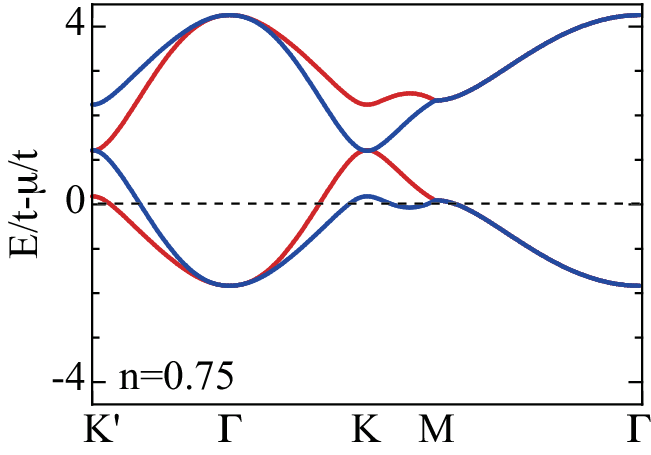} &        
        \includegraphics[width=0.18\linewidth]{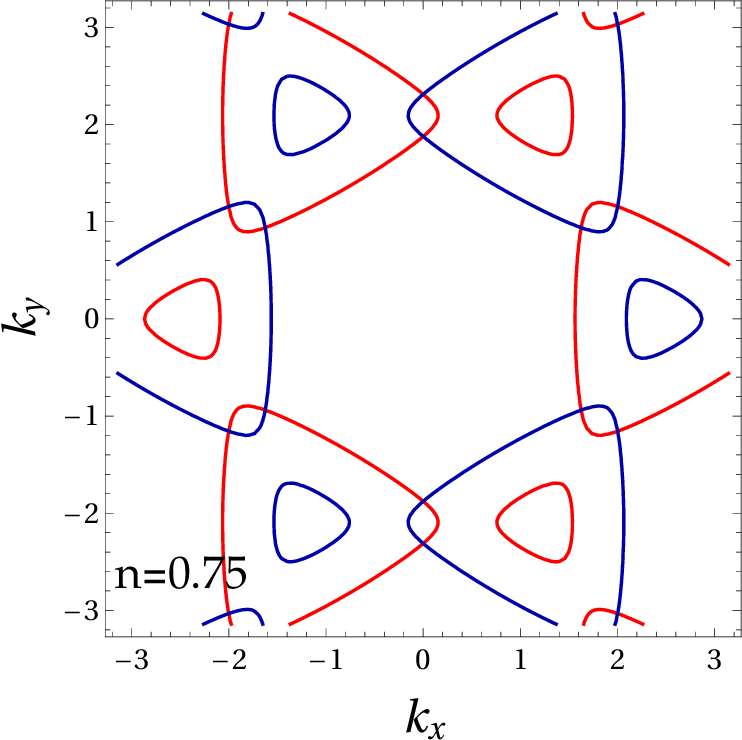} \\ \\
    \end{tabular}
         \begin{tabular}{cccc}
            (f) & & (g) \\ 
        \includegraphics[width=0.25\linewidth]{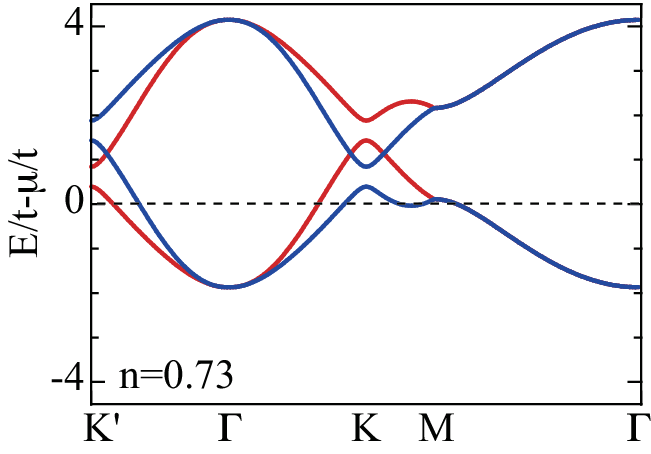} & 
        \includegraphics[width=0.18\linewidth]{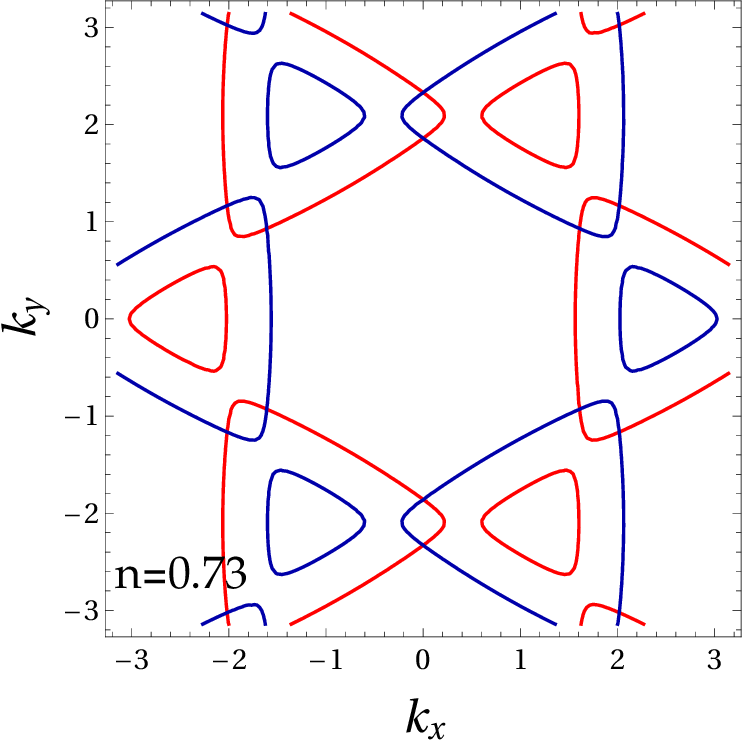}&~~~~
        \includegraphics[width=0.25\linewidth]{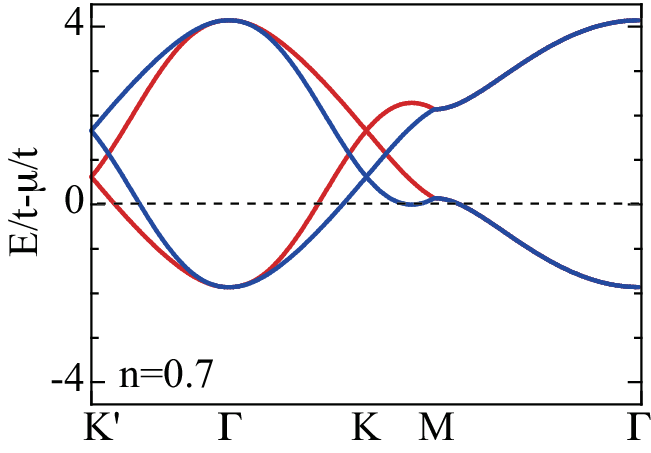} &        
        \includegraphics[width=0.18\linewidth]{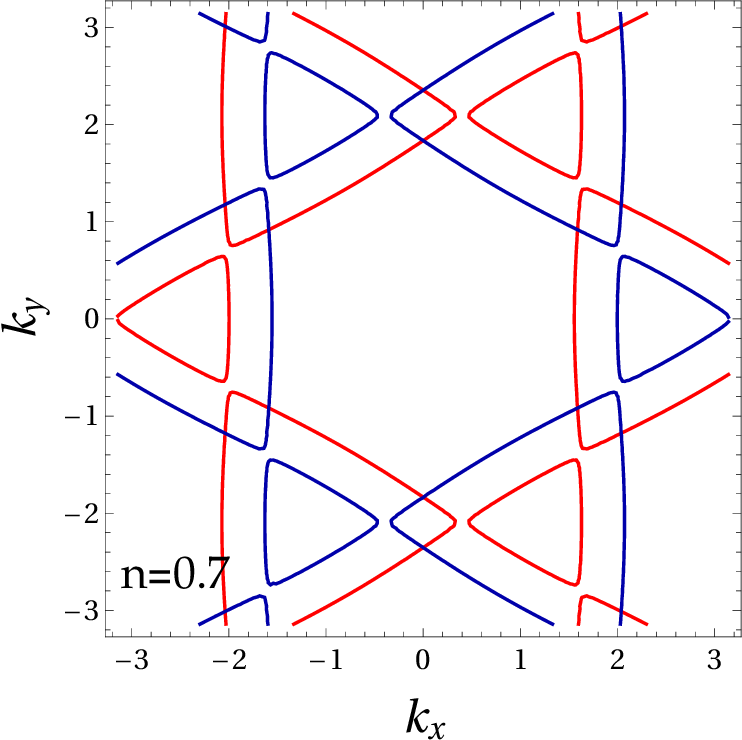} 
    \end{tabular}        
\caption{Mean-field results for $\lambda= 0.1t$ and $J_z=2t$.
(a) Order parameters and chemical potential as a function of electron density $n$ and band structures for $n=1$. Inset:  the first Brillouin zone is depicted and the black lines indicate the scans considered here. Band structures and corresponding Fermi surfaces for (b) $n=0.9$, (c) $n=0.8$, (d) $n=0.79$, (e) $n=0.75$, (f) $n=0.73$, and (g) $n=0.7$. The red (blue) line corresponds to the spin up (down) component.}
\label{fig:MF}
\end{figure*}

\begin{figure*}
    \centering
    \begin{tabular}{c}
            (a) \\
        \includegraphics[width=0.5\linewidth]{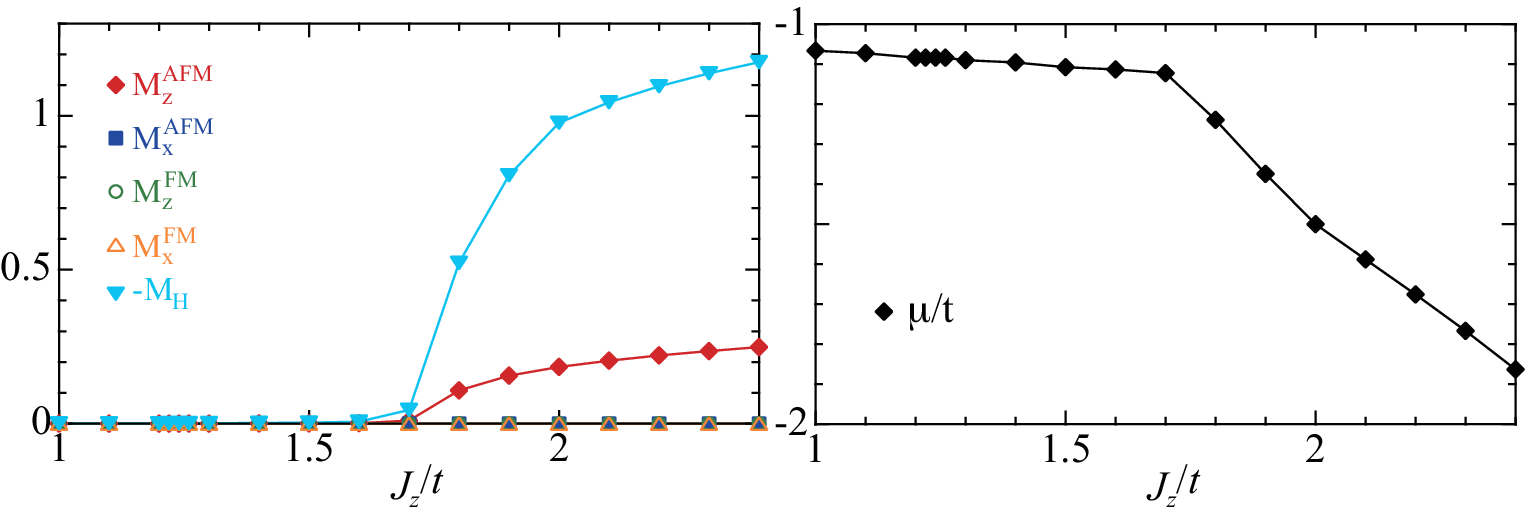} \\ 
    \end{tabular}
     \begin{tabular}{cc}
            (b) &  (c) \\ 
        \includegraphics[width=0.45\linewidth]{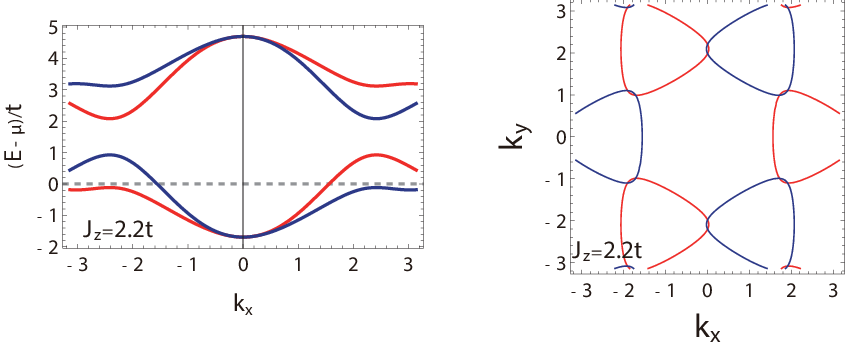} & 
        \includegraphics[width=0.45\linewidth]{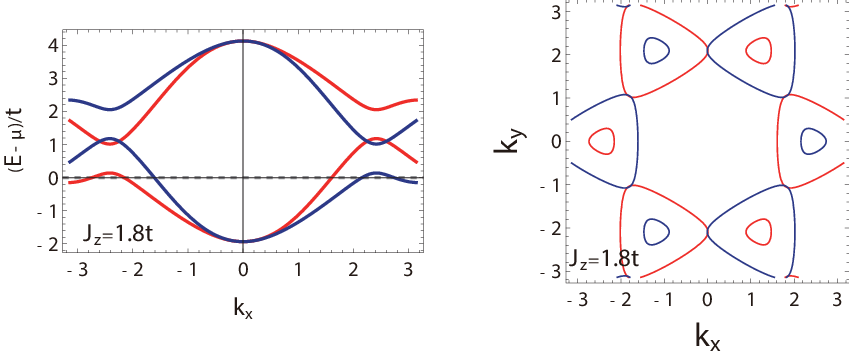} \\ 
    \end{tabular}
     \begin{tabular}{cccc}
            (d) & (e) \\ 
        \includegraphics[width=0.45\linewidth]{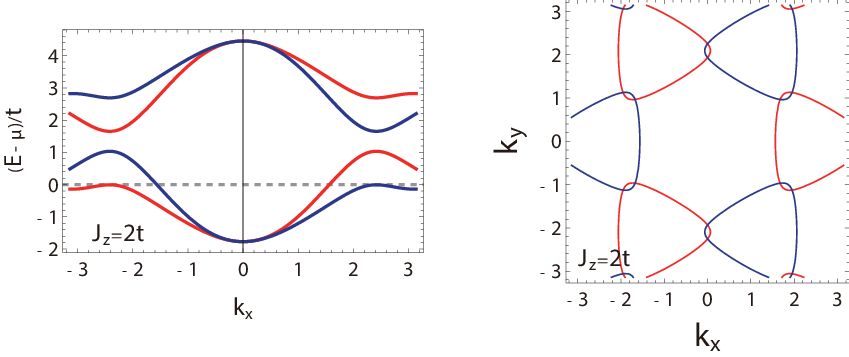} & 
        \includegraphics[width=0.45\linewidth]{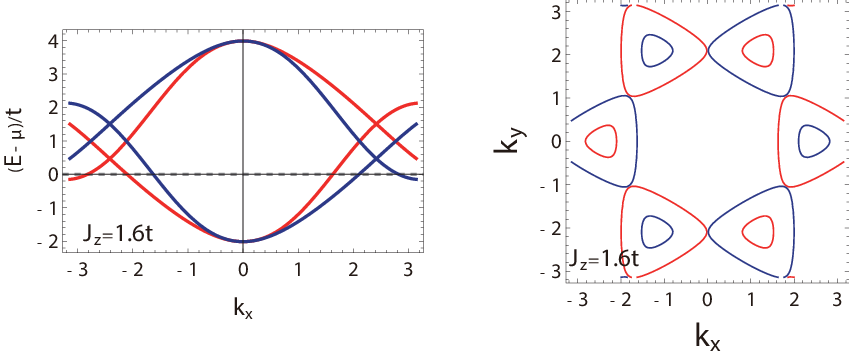} \\ 
    \end{tabular}    
\caption{
(a) Order parameters and chemical potential as a function of $J_z$ for $\lambda=0.1t$ and $n=0.79$. 
Band structures and corresponding Fermi surfaces along $k_y=0$ for (b) $J_z=2.2t$, (c) $J_z=2t$, (d) $J_z=1.8t$, (e) $J_z=1.6t$.
 The red (blue) line corresponds to the spin up (down) component.
}
\label{Jzdep-n}
\end{figure*}

\begin{figure*}
\centering
     \begin{tabular}{cccc}
            (a) & (b) & (c) & (d)\\ 
         \includegraphics[width=0.22\linewidth]{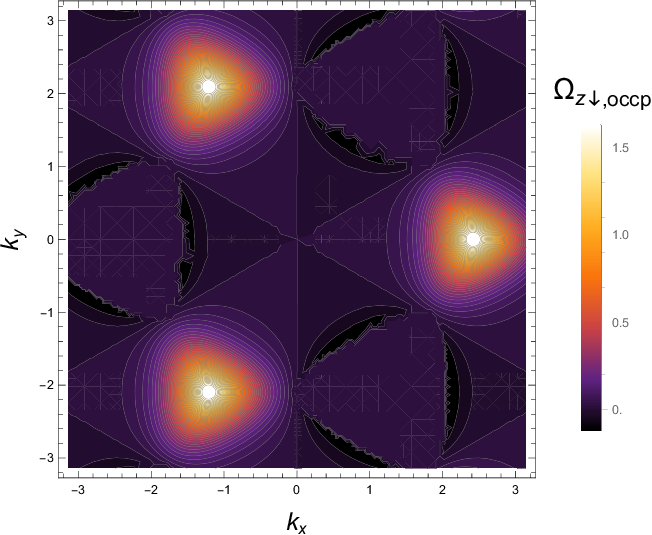} & 
         \includegraphics[width=0.22\linewidth]{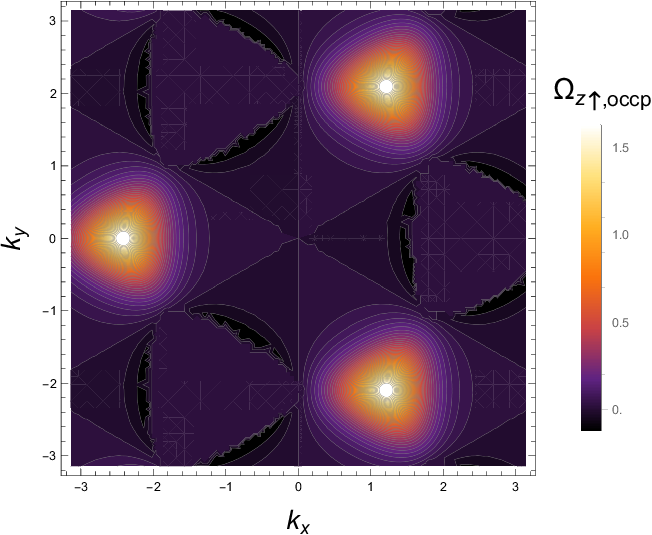} & 
         \includegraphics[width=0.22\linewidth]{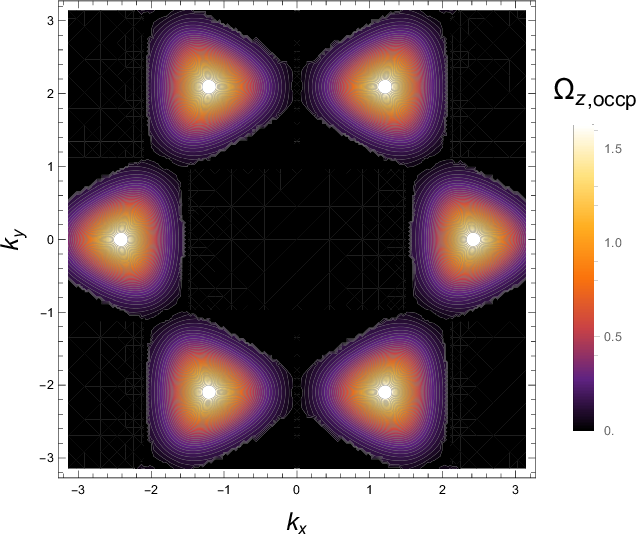}&  
        \includegraphics[width=0.24\linewidth]{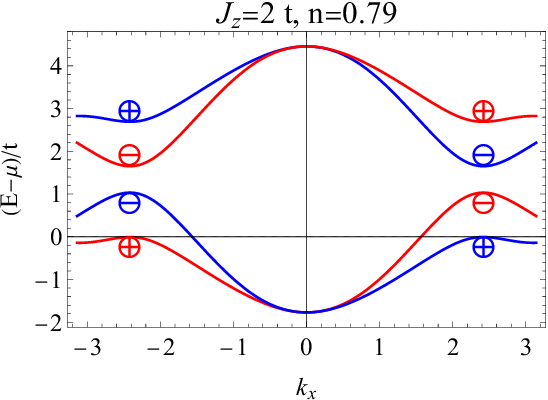}
    \end{tabular}
\caption{
Contour plots of the Berry curvatures (BC) of the occupied spin sub-bands with a spin projection (a) $\sigma=-1$ and (b) $\sigma=1$.
The total Berry curvature of the occupied bands is depicted in (c).
The corresponding band structures are shown in (d)  where the symbols $\oplus$ and $\ominus$ indicate the sign of the BC around the Dirac valleys.
Calculations are done for $J_z=2t$, $n=0.79$.
The BC is in unit of $a^2$, where $a$ is the distance between NN atoms.
}
\label{BC}
\end{figure*}

The doping dependence of the corresponding anomalous Hall conductivity is depicted in Fig.~3 in the main text. This behavior is reminiscent of that of the secondary order parameter denoted $M_{\rm H}$ (= $O_{\rm AH}$ given in the main text) in Fig.~\ref{fig:MF}.
As shown in Fig.~3 of the main text, by decreasing $n$ from half-filling, $|\sigma_{\rm Hall}|$ increases up to a maximum value corresponding at a doping amplitude $n_{c}$ below which it drops and vanishes at a critical doping value $n_{c0}$. 
The increase of $|\sigma_{\rm Hall}|$ can be understood from the BC contribution which is enhanced as the area of the spin-polarized Fermi surface increases [see Fig.~\ref{fig:MF} for $0.79<n<1$]. 
The drop of $|\sigma_{\rm Hall}|$ below $n_{c}$ is due to two effects: (i) the sharp decrease of $M^{\text{AFM}}_z$ [see Table~\ref{Mz}] and (ii) the contribution to the BC of two bands with opposite spin projections at a given valley [see Fig.~\ref{fig:MF} for $n<0.79$]. In the case of Fig.~\ref{fig:MF}, the larger the Fermi surface of the conduction band of spin $\sigma=-$ at the valley $K$, the smaller the BC.

\begin{figure*}
\centering
\centerline{\includegraphics[width=0.76\linewidth]{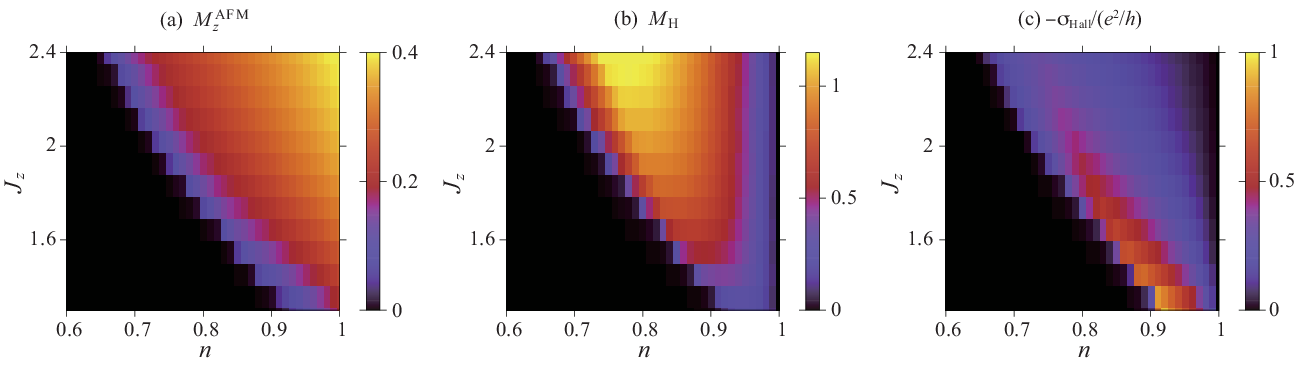}}    
\caption{
Order parameters and anomalous Hall conductivity in the electron density $n$ versus the interaction $J_z$ for $\lambda= 0.1t$.}
\label{n-Jz}
\end{figure*}

Figure~\ref{n-Jz} shows the coupling and the doping ranges where emerge the order parameters in a non-vanishing anomalous Hall conductivity.

\subsection{Valley-Zeeman spin-orbit coupling}
In the continuum limit, our modified Kane-Mele term for the Hamiltonian of Eq.~(1) in the main text, $a_{\mathbf{k}}\tau^0\sigma^z$, reduces around a Dirac point, $\mathbf{k}=\mu_z \mathbf{K}$, to $-3\sqrt{3}\lambda \mu_z \sigma^z\tau^0$, where $\mu_z=\pm$ indicates the valley index.
It is interesting to note that the resulting expression corresponds to a valley-Zeeman spin-orbit coupling (VZ-SOC), as observed in transition metal dichalcogenides (TMDs)~\cite{Xiao12,Wang2022}.
The VZ-SOC breaks inversion symmetry and induces, in monolayer TMD, a spin-split electronic band structure and a spin-valley locking effect, which gives rise to the valley anomalous Hall effect~\cite{Xiao12,Chen17}.
While these phenomena are similar, their origins differ: in our case, the modified Kane-Mele term is characterized by a spin-dependent hopping process, while in TMD, it arises from the band structure formed by the coupling of transition metal $d$-orbitals and chalcogen $p$-orbitals~\cite{Zhu2011}, further influenced by atomic spin-orbit coupling that splits the bands.

\subsection{Effect of Rashba spin-orbit coupling and repulsive Hubbard interaction}
\begin{figure*}
\centering
     \begin{tabular}{cc}
            (a) & ~~~~~~~~  (b) \\ 
       \includegraphics[width=0.28\linewidth]{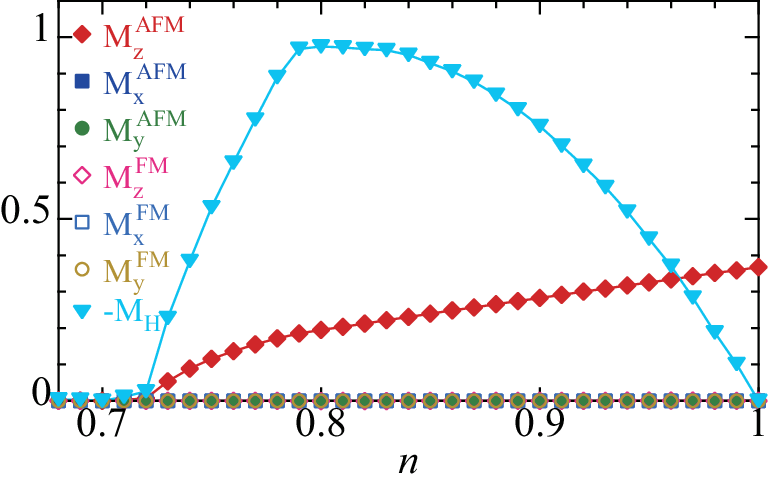} & ~~~~~~~~
       \includegraphics[width=0.28\linewidth]{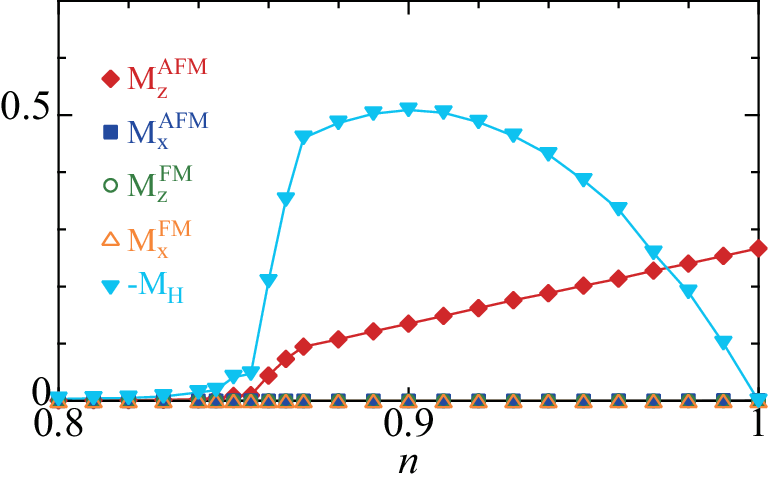} \\ 
    \end{tabular}
\caption{
Mean-field results with additional terms in Eq. (2) of the main text: (a) Rashba spin-orbit coupling [$\lambda_R = 0.1t$ at $J_z = 2t$] and (b) Hubbard $U$ term [$U = 2.4t$ at $J_z = 0.4t$]. Here, $\lambda = 0.1t$.
}
\label{realJz-Rashba}
\end{figure*}

We present the mean-field results when an additional term is included in Eq.(2) of the main text.
In Fig.~\ref{realJz-Rashba}(a) we show the results in the presence of a Rashba SOC which does not affect the altermagnetic properties of the system. 
Including the repulsive Hubbard $U$ term demonstrates that the altermagnetism persists even when the coupling $J_z$ is smaller than $U$ and $t$, as shown in Fig.~\ref{realJz-Rashba}(b).

\begin{table*} 
\begin{tabular}{|c|c|c|c|c|c|c|c|}
\hline
     $n$&  1&0.9&0.8& 0.79&0.75&0.73&0.7\\
\hline     
    $M^{\text{AFM}}_z$ & 0.3675933&0.2839518&0.1946018&0.1843236& 0.1134823& 0.0497554&0.0015749\\
    \hline
\end{tabular}
\caption{\label{Mz}Mean-field data of $M^{\text{AFM}}_z$ at different doping levels $n$ for $J_z=2t$.}
\end{table*}

\end{document}